\documentclass[prd,preprint,superscriptaddress]{revtex4}
\usepackage{revsymb}
\usepackage{graphicx}

\newcommand{\be}{\begin{equation}}
\newcommand{\ee}{\end{equation}}
\newcommand{\ben}{\begin{eqnarray}}
\newcommand{\een}{\end{eqnarray}}
\newcommand{\n}{\label}
\newcommand{\no}{\noindent}

\begin{document}

\title{Interacting quintessence solution to the coincidence problem}

\author{Luis P. Chimento}
\affiliation{Departamento de F\'{\i}sica,
Facultad de Ciencias Exactas y Naturales,
Universidad de Buenos Aires,
Ciudad  Universitaria,  Pabell\'on  I,
1428 Buenos Aires, Argentina.}
\author{Alejandro S. Jakubi}
\affiliation{Departamento de F\'{\i}sica,
Facultad de Ciencias Exactas y Naturales,
Universidad de Buenos Aires,
Ciudad  Universitaria,  Pabell\'on  I,
1428 Buenos Aires, Argentina.}
\author{Diego Pav\'on}
\affiliation{Departmento de F\'{\i}sica, Facultad de Ciencias,
Edificio Cc, Universidad Aut\'onoma de Barcelona. E-08193
Bellaterra (Barcelona). Espa\~{n}a.}
\author{Winfried Zimdahl}
\affiliation{Fachbereich Physik,
Universit\"{a}t Konstanz, PF M678,
D-78547, Konstanz, Deutschland.}

\begin{abstract}
We show that a suitable interaction between a scalar field and a matter
fluid in a spatially homogeneous and isotropic spacetime can drive the
transition from a matter dominated era to an accelerated expansion
phase and simultaneously solve the coincidence problem of our present
Universe. For this purpose we study the evolution of the energy
density ratio of these two components. We demonstrate that a
stationary attractor solution is compatible with an accelerated
expansion of the Universe. We extend this study to account for
dissipation effects due to interactions in the dark matter fluid.
Finally, Type Ia supernovae and primordial nucleosynthesis data are
used to constrain the parameters of the model.
\end{abstract}


\maketitle

\section{Introduction}

Nowadays there is a wide consensus among observational cosmologists that our
Universe is accelerating its expansion -for a pedagogical short update see
\cite{pedagogical}, see also \cite{
Perlmutter98,Riess98,paris,Efstat,Perlmutter99,Bahcall99,Efstat02}- which
implies that the Einstein-de Sitter scenario has to be abandoned, at least
to describe the present era.
The nature of the dark energy behind this acceleration is unknown. Two main
proposals, namely the $\Lambda$CDM model and QCDM models have been advanced.
The former assumes a cosmological constant arising from the energy density of
the zero point fluctuations of the quantum vacuum and cold dark matter in the
form of pressureless dust. While it fits rather well all the observational
constraints \cite{melchiorri} it has serious difficulties with the low
observed value
of this vacuum energy (that by all accounts should be many orders of
magnitude
higher) and fails to address the so--called ``coincidence problem", namely why
the energy density of both components happen to be of the same order today?
The second group of models assume an evolving scalar field possessing a
negative pressure and cold dark matter. It also fits well the observational
constraints, and seems rather natural but it is not clear whether it really
solves the coincidence problem (for a recent review see \cite{paris}).

Most of the QCDM models assume that the dark matter and the scalar
field components evolve independently. However, given that the
physical nature of the quintessence field is still unknown and also
that the dark matter may well be a substratum not as simple as a
pressureless perfect fluid, there seems to be no a priori reasons to
exclude a coupling between both components.  Interacting quintessence
models have been shown to provide qualitatively new features which
may be relevant to the coincidence problem
\cite{luca,zpc}. In particular, it has been demonstrated that a
suitable coupling may give rise to a stable constant ratio of the
energy densities of both components which is compatible with an
accelerated expansion of the Universe \cite{zpc}.  On the other hand,
this model could not answer the question of how such a stationary
solution can be obtained as the result of a dynamical evolution.  In
the present paper we clarify this point and establish an exactly
solvable model for a smooth transition from a matter dominated phase
to a subsequent period of accelerated expansion. This model implies
the evolution of the density ratio towards a finite, stable,
asymptotic value. Thus it may represent a solution to the coincidence
problem.

As shown in another previous paper, a very suitable ingredient of quintessence
models is a dissipative, negative, scalar pressure of the matter component.
Such a quantity may simultaneously help to drive acceleration and solve the
coincidence problem \cite{enlarged}. A negative pressure arises naturally
from bulk viscous dissipation, quantum particle production or self-interaction
in the matter component \cite{antf}. Here we combine the advantages of
quintessence models interacting with matter (QIM) with those relying on
a dissipative pressure within the latter.

The aim of this paper is to show, that on this basis a solution of the coincidence
problem in an accelerating universe can be realized in a comparatively simple
manner within the framework of general relativity.

The paper is organized as follows. Section II introduces the basic equations
of the model. Section III explores the dynamics of the energy density ratio,
including the stability properties of the stationary solutions. Furthermore,
it derives the corresponding scalar field potential. Section IV investigates
the role of a dissipative pressure within the dark matter component and
discusses the behavior of the deceleration parameter. In section V the
available magnitude--redshift data of supernovae Type Ia (SNe Ia) are used in
combination with primordial nucleosynthesis data to restrict the parameters of
the model. Section VI presents our conclusions and final comments. Lastly, the
Appendix discusses briefly the connections of the matter-quintessence coupling
with cosmological inhomogeneities, the issue of possible anomalous
acceleration of baryonic matter and some consequences of this interaction on
the early universe. Units have been chosen so that $c = 8\pi \, G = 1$.

\section{Interacting cosmology}

Let us assume a FLRW spacetime with matter and a minimally coupled scalar
field. The Friedmann equation and the overall conservation equation read

\be
3 H^{2} + 3 \frac{k}{a^{2}} =  \rho_{m} + \rho_{\phi}
\qquad (k = 1, 0, -1),
\label{feq}
\ee

\noindent
and

\be
\dot{\rho}_{m}+\dot\rho_{\phi}+3H(\gamma_m\rho_m+\gamma_{\phi}\rho_\phi)=0 ,
\label{ceq}
\ee

\noindent
respectively, where have assumed the  equations of state $p_{m} =
(\gamma_{m} - 1) \rho_{m}$, $p_{\phi} = (\gamma_{\phi} - 1)
\rho_{\phi}$ with $ 1 \leq \gamma_{m} \leq 2$ and $0 \leq
\gamma_{\phi} \leq 2$.
We also introduce an overall effective baryotropic index by

\be
\gamma \rho = \gamma_{m} \, \rho_{m} +  \gamma_{\phi} \, \rho_{\phi},
\label{overall}
\ee

\noindent
where $\rho = \rho_{m} + \rho_{\phi}$ is the total energy
density ($\rho_{m}$ is assumed to include both baryonic and nonbaryonic
matter; see the Appendix for a discussion of this point). Then Eqs.
(\ref{feq}) and (\ref{ceq}) can be written as

\be
3 \left(H^{2} + \frac{k}{a^{2}}\right) = \rho \, ,
\label{f1eq}
\ee

\noindent
and

\be
\dot{\rho} + 3\gamma H \rho = 0.
\label{c1eq}
\ee

\no
respectively.
In terms of the density parameters $\Omega_{m} \equiv \rho_{m}/(3H^{2})$,
$\Omega_{\phi} \equiv \rho_{\phi}/(3 H^{2})$ and $\Omega_{k} = k/(aH)^{2}$,
the last two equations become

\be
\Omega_{m} + \Omega_{\phi} + \Omega_{k} = 1 \, ,
\label{6}
\ee
\\
and
\\
\be
\dot{\Omega} = \Omega (\Omega - 1) (3\gamma -2) H \, ,
\label{7}
\ee
\\
where $\Omega \equiv \Omega_{m} + \Omega_{\phi}$.
This scheme is compatible with an interaction between the scalar
field and the matter, described by a coupling term $\delta$ according
to
\\
\be
\dot{\rho}_{m} + 3H \gamma_m\rho_{m} = \delta \ ,
\label{deltam}
\ee
\\
and
\\
\be
\dot{\rho}_{\phi} + 3H \gamma_\phi\rho_\phi = -\delta\ .
\label{deltaphi}
\ee
\\
The coupling is left unspecified at this stage. It represents an
additional degree of freedom which will be used below to guarantee
the existence of solutions with a stationary energy density ratio.
After introducing a generalized dissipative pressure through
$\delta \equiv -3\Pi H$, the last two equations take the form
\\
\be
\dot{\Omega}_{m} + 3H\left(\frac{2 \dot{H}}{3H^{2}} +
\gamma_{m}^{(e)}\right) \Omega_{m} = 0,
\label{domegam}
\ee
\\
\be
\dot{\Omega}_{\phi} + 3H\left(\frac{2\dot{H}}{3H^{2}} +
\gamma_{\phi}^{(e)}\right) \Omega_{\phi} = 0,
\label{domegaphi}
\ee
\\
respectively, where we have introduced the effective baryotropic indices
\\
\be
\gamma_{m}^{(e)} = \gamma_{m} + \frac{\Pi}{\rho_{m}}, \qquad
\gamma_{\phi}^{(e)} = \gamma_{\phi} - \frac{\Pi}{\rho_{\phi}}.
\label{effectivep}
\ee
\\
\noindent
{}From Eqs. (\ref{domegam}) and (\ref{domegaphi}) it follows that the
energy density ratio $r \equiv \rho_m /\rho_\phi = \Omega_m/\Omega_\phi$ obeys
the equation
\\
\be
\n{dr}
\dot r= -3Hr\left[\gamma_m^{(e)}-\gamma_\phi^{(e)}\right].
\ee
\\
It describes the dynamics of the parameter $r$ in terms of the equations
of state and the mutual interaction of the components. Formally it looks as if
we were dealing with a noninteracting dissipative matter fluid
(cf. \cite{enlarged,qsa}).

We look for a dynamical solution to the coincidence problem such that the
Universe approaches a stationary stage in which $r$ becomes a constant.
A nonvanishing constant solution to Eq. (\ref{dr}) occurs when the
stationary condition

\be
\gamma_{m}^{\left(e\right)} =
\gamma_{\phi}^{\left(e\right)} \ ,
\label{rstat}
\ee

\noindent
holds. In virtue of (\ref{rstat}) the overall baryotropic index on
a stationary solution (subindex $s$) is given by

\be
\gamma_{s} = - \frac{2 \dot{H}}{3H^{2}} =
\gamma_{\phi} - \frac{\Pi}{\rho_{\phi}} =
\frac{1}{1+r}\left[\left(\gamma_{\phi} - \frac{\Pi \rho}
{\rho_{m} \rho_{\phi}}\right) r + \gamma_{\phi}\right],
\label{gstat}
\ee

Indeed, the simplest solution to the the cosmic coincidence problem occurs
when $\Omega_m = \Omega_{m s}$ and $\Omega_{\phi} = \Omega_{\phi s}$, with
$\Omega_{m s}$ and $\Omega_{\phi s}$ constants. We call this case  ``strong
coincidence". Then, using Eqs (\ref{domegam}) and (\ref{domegaphi}), the
stationary conditions become

\be
\gamma_{m}^{\left(e\right)} =
\gamma_{\phi}^{\left(e\right)} =
-\frac{2\dot H}{3H^2}.
\label{stationary}
\ee

\noindent
Combining Eqs. (\ref{f1eq}), (\ref{c1eq}) and (\ref{gstat}) it follows that
on the stationary solution

\be
\frac{k}{a^{2}} = - k \frac{\dot{H}}{(Ha)^{2}}.
\label{simply}
\ee

\noindent
This last equation becomes an identity for $k = 0$ while for $k \neq
0$ it leads to $a(t) \propto t$. This second possibility implies a
nonaccelerating universe. This means that under the strong coincidence
condition an accelerated expansion is only possible in a flat FLRW
universe. Then Friedmann's equation reduces to $3 H^{2} = \rho$,
thus, $\Omega = 1$ and $\dot{\Omega}_m = -\dot{\Omega}_\phi$.
Further, assuming $\gamma_{s} \simeq 2/(3 \alpha)$, with $\alpha>1$ a
constant, it follows from Eq. (\ref{gstat}) that $a(t) \propto
t^{\alpha}$ and $\rho_s\propto t^{-2}$. Of course, this does not
indicate how such solution is approached.

\section{Dynamics and stability}
The purpose of this section is a detailed study of the general dynamics of the
density ratio $r$ as given by Eq. (\ref{dr}). At first we look for constant
solutions $r=r_s$, representing a stationary stage of the Universe. We will
assume that all the quantities in
$\Gamma\equiv\gamma_m^{\left(e\right)}-\gamma_\phi^{\left(e\right)}$ can be
expressed in terms of the ratio $r$. Then, according to Eq. (\ref{rstat}),
stationarity requires $\Gamma(r_s)=0$. Let us look at the stability of these
constant solutions. Expanding the general solution of Eq. (\ref{dr}) about
$r_s$ in powers of $\epsilon\equiv r-r_s$, we get up to
first order in $\epsilon$
\\
\be
\n{dep}
\dot\epsilon=-3Hr_s\left[\frac{d\Gamma}{dr}\right]_{r=r_s}\epsilon.
\ee
\\
\noindent
Eq. (\ref{dep}) shows that a root $r_s$ of $\Gamma(r_s)=0$ is
asymptotically stable solution whenever $\left[d\Gamma/dr\right]_{r=r_s}>0$.
Identifying the energy-momentum tensor of the scalar field $\phi$ with that of
a perfect fluid

\be
\n{rr}
\rho_\phi=\frac{1}{2} \dot\phi^2 + V(\phi), \qquad
p_\phi=\frac{1}{2} \dot\phi^2 - V(\phi),
\ee

\noindent
and using the Eq. (\ref{deltaphi}), we get

\be
\n{p/rm}
\frac{\Pi}{\rho_\phi}=\gamma_\phi-r\Gamma
\left[
\frac{1}{2-\gamma_\phi}\frac{d\gamma_\phi}{dr}+\frac{1}{V}\frac{dV}{dr}\right].
\ee

\noindent
Likewise, using the relation

\be
\n{rv}
\rho_\phi=\frac{2V}{2-\gamma_\phi},
\ee

\noindent
the Friedmann equation can be recast as

\be
\n{00r}
3H^2=\frac{2(1+r)V}{2-\gamma_\phi}-3\frac{k}{a^{2}}.
\ee

The equations (\ref{dr}), (\ref{p/rm}) and (\ref{00r}) provide the following
solution procedure. In a first step we specify  the indices $\gamma_m$ and
$\gamma_\phi$ and the ratio $\Pi/\rho_\phi$ as functions of $r$. With Eqs.
(\ref{rstat}) and (\ref{dep}) we may calculate the constant solutions $r_s$ as
the roots of $\Gamma(r_s)=0$ and check their stability properties. In the
second step we integrate the Eq. (\ref{dr}) to obtain $r=r(a)$. This provides
us with the dynamics of the density ratio that is relevant for the solution of
the coincidence problem. Constraints from nucleosynthesis, CMB anisotropy and
cosmic structure formation preclude an early quintessence dominance stage
\cite{pedagogical,paris}. We will also assume that the current density ratio
$r_0\simeq 0,56\pm 0.07$ \cite{flat,tytler,turner2001} is close to a constant
attractor solution. As the coincidence problem may be phrased in terms of the
``why now'' question, the dynamical solution to this problem arises because
the variation of the density ratio $r$ is quite small so that there is nothing
very peculiar about the present time and the value $r_0$. Hence, we shall seek
to describe the transition from an matter dominance with $r\gg 1$ to a stable
stationary era with $r\alt 1$ (coincidence era).

Furthermore, from Eq. (\ref{p/rm}) it is possible to obtain $V=V(r)$ and from
Eq. (\ref{00r}) one finds $a=a(t)$. In addition, the relation between the
kinetic energy of the scalar field and its potential

\be
\n{k}
\dot\phi^2=\frac{2\gamma_\phi}{2-\gamma_\phi}\,V\,,
\ee

\noindent
can be integrated to give $\phi=\phi(t)$ and this function inverted to yield
the potential by $V(\phi)=V(a(t(\phi)))$.

We shall apply the indicated procedure to an interaction
characterized by $\Pi=-c^2\rho$ with $c^2$ a constant and $k=0$ which
has already been discussed in \cite{zpc}. (A more detailed discussion
of the corresponding coupling  $\delta = 3 c^{2}\rho H$ is given in the
Appendix). When $\gamma_m$ and $\gamma_\phi$ are assumed to be constants,
the stationary solutions of (\ref{dr}) are obtained by solving
$r_s\Gamma(r_s)=0$. The roots of this  quadratic equation are
\\
\be
\n{r0pm}
r_s^\pm=-1+\frac{\gamma_m-\gamma_\phi}{2c^2}\pm\sqrt\Delta,
\ee
\\
\noindent
where the discriminant
\\
\be
\n{del}
\Delta\equiv\frac{\gamma_m-\gamma_\phi}{c^2}\left[\frac{\gamma_m-\gamma_\phi}{
4c^2} -1\right]
\ee
\\
\no
must be nonnegative to obtain real solutions $r_s^\pm$. The quantity
$\Delta$ determines the difference between the stationary values $r_s^+ -
r_s^- = 2\sqrt\Delta$, and the relationship $r_s^+  r_s^- = 1$ holds,
implying $r_s^+ \geq 1 \geq r_s^-$. It is expedient to write
Eq. (\ref{dr}) in terms of $r_s^-$ and $r_s^+$
\\
\be
\dot r=-3c^2H\left[(r-r_s^-)(r_s^+ -r)\right]\, .
\label{dr1}
\ee
\\
\noindent
When $\Delta>0$, the stability of these solutions is
determined by the sign of
\\
\begin{equation} \label{sgn}
\left.\frac{\partial \dot r}{\partial r}\right|_{r_s^\pm}
=\pm 3c^2H \left(r_s^+-r_s^-\right) \, .
\end{equation}

\noindent While $r_s^+$ is unstable, the solution $r_s^-$ is asymptotically
stable. This means that a solution of Eq. (\ref{dr1}) starting at $r_s^+$ and
and decreasing towards $r_s^-$ fits in the above picture regarding the
evolution of the density ratio. In this picture $r_{s}^{+}$ stands for the
density ratio at the onset of the quintessence--matter interaction. On the
other hand, when $\Delta = 0$, corresponding to the quadratic root
$r_s^+ = r_s^- = 1$, the density ratio is growing so that we will not consider
it any further.

The family of regular monotonic decreasing solutions of Eq. (\ref{dr1})
in the range $r_s^+ > r > r_s^-$ is given by

\be
r(x)=\frac{r_s^- + xr_s^+}{1+x} \, ,
\label{rg}
\ee

\no where $x=(a/a^*)^{-\lambda}$, $\lambda\equiv 6c^2\sqrt\Delta$ and
$r^*\equiv r(1) = \left(r_s^+ + r_s^-\right)/2$. In the following we will
denote by an asterisk magnitudes at the epoch of mean density ratio $r=r^*$, or
equivalently $x=1$. For $a \ll a^*$, corresponding to $x \gg 1$, we have $r
\approx r_s^+$, while in the opposite case $x \ll 1$ the stable solution $r
= r_s^-$ is approached. Two other families of solutions of Eq.
(\ref{dr1}) exist for $r<r_s^-$ and $r>r_s^+$. As they are singular and
exhibit a growing ratio, they will not be considered here.

Rewriting Eq. (\ref{p/rm}) in terms of the variable $r$, we have
\\
\be
\n{V1}
\int \frac{dV}{V}
 =-\frac{1}{c^2}\int \frac{\gamma_\phi+c^2(1+r)}{(r-r_s^-)
(r-r_s^+)}\,dr .
\ee
\\
Integrating (\ref{V1}), we obtain the history of the
quintessence potential after some algebra and using
$r\Gamma=4c^2x\Delta/(1+x)^2$
\\
\be
\n{po}
V(x)=\frac{1}{2}V^* \left[1 + x\right]
x^{3\gamma_\phi^-/\lambda} \ ,
\ee
\\
where $\gamma_\phi^\pm=\gamma_\phi+c^2(1+r_s^\pm)$.
The expressions (\ref{rg}) for $r$ and (\ref{po}) for $V$ determine the Hubble
rate according to (\ref{00r}).
In terms of the redshift $z= a_0/a -1$
the latter becomes (for a spatially flat universe)
\\
\begin{equation}
H = H_0\left[\frac{1 + \sigma \left(1+z\right)^\lambda}{
1 + \sigma }\right]^{1/2} \left(1+z\right)^{\frac{3}{2}\gamma_\phi^-}
\ .
\label{Hz}
\end{equation}
\\
Here we have introduced the quantity

\begin{equation}
\sigma \equiv \frac{1 + r_s^+}{1 + r_s^-}
\left(\frac{1}{1+z^*}\right)^\lambda=
\frac{1 + r_s^+}{1 + r_s^-}\frac{r_0-r_s^-}{r_s^+-r_0}
\ ,
\label{sigma}
\end{equation}

\noindent
and used the transformation
\\
\begin{equation} \label{xz}
x=\left(\frac{1+z}{1+z^*}\right)^\lambda .
\end{equation}
\\
This parameter $\sigma$ is a measure of the closeness of the present
Universe to the asymptotic attractor (stationary) stage, as $\sigma=0$
corresponds to the constant solution $r=r_s^-$.

With the help of (\ref{rg}), the equations for the energy densities of the
matter (\ref{deltam}) and the field (\ref{deltaphi}) can be integrated, which
results in

\be
\n{rfg}
\rho_\phi=\frac{1}{2}\rho^*(1+x)x^{3\gamma_\phi^-/\lambda}\,, \qquad
\rho_m=\frac{1}{2}\rho^*(r_s^- + xr_s^+)x^{3\gamma_\phi^-/\lambda}\,
\ee

\noindent where the constants are related by $2V^*=\rho^*(2-\gamma_\phi)$ .
Using Eqs. (\ref{00r}), (\ref{rg}) and (\ref{po}) we integrate $\dot
x/x=-\lambda H$ to obtain the scale factor $a(t)$ in an implicit form in terms
of the hypergeometric function

\begin{equation} \label{tx}
t=\frac{2}{\gamma_\phi^-}
\left[\frac{2-\gamma_\phi}{3V^*\left(r_s^-+1\right)}\right]^{1/2}x^{-D/2}
{}_2F_1\left(\frac{1}{2},-\frac{D}{2},-\frac{D}{2}+1;-\frac{x}{B}\right)
\end{equation}

\noindent
where $B=\left(r_s^-+1\right)/\left(r_s^++1\right)$ and
$D=3\gamma_\phi^-/\lambda$. Similarly we can integrate Eq. (\ref{k}) to obtain
the scalar field

$$
\phi(x)-\phi^*=\frac{1}{\lambda}\left(\frac{3\gamma_\phi}{r_s^++1}\right)^{1/2}
\left[
\ln  \left( {\frac {B+3+2\,\sqrt {2(B+1)}}{B+1+2\,x+2\,\sqrt {x
+1}\sqrt {x+B}}} \right)\right.
$$

\begin{equation} \label{phix}
+\left.{\frac {1}{\sqrt {B}}\ln  \left( {\frac {2\,B+xB+x+2\,\sqrt {B}
\sqrt {x+1}\sqrt {x+B}}{x \left( 3\,B+1+2\,\sqrt {2B}\sqrt {B+
1} \right) }} \right) }
\right]
\end{equation}

\noindent and combined with Eq. (\ref{po}) it yields the potential $V(\phi)$
in parametric form. As both $V(x)$ and $\phi(x)$ are monotonic functions, we
find that $V(\phi)$ is also monotonic. Finally, combining Eq. (\ref{phix})
with (\ref{tx}) we obtain $\phi(t)$ in implicit form.

In the near attractor regime simple, explicit expressions arise. For $r
\approx r_s^-$ the history of the potential (\ref{po}) can be approximated by
$V\simeq (1/2)V^*(a/ a^*)^{-3\gamma_\phi^-}$ and Eq. (\ref{00r}) becomes

\be
\n{00a}
3H^2\simeq\frac{2(1+r_s^-) V^*}{2-\gamma_\phi}
\left(\frac{a^*}{a}\right)^{3\gamma_\phi^-}.
\ee

\noindent
Hence the evolution in this regime is near power--law
\\
\be
\n{aa}
a(t) \simeq  a^*\, \left(\frac{t}{t^*}\right)^{2/3\gamma_\phi^-}\,,
\qquad
V\simeq \frac{1}{2} V^*\left(\frac{t^*}{t}\right)^2\,.
\ee
\\
We also have the approximate expressions
\\
\be
\n{rpat}
\rho_\phi\simeq \frac{2 V^*}{2-\gamma_\phi}
\left(\frac{a^*}{a}\right)^{-3\gamma_\phi^-}\,, \qquad
\rho_m\simeq r_s^-\rho_\phi, \qquad
\phi\simeq \sqrt{\frac{2\gamma_\phi V^* {t^*}^2}
{2-\gamma_\phi}}\ln\frac{t}{t^*}+\phi^*,
\ee
\\
where $\phi^*=\phi(t^*)$ and the consistency relation
\\
\be
\n{con}
V^*=\frac{3(2-\gamma_\phi){H^*}^2}
{1+r_s^-}
\ee
\\
holds with $H^*\equiv 2/(3\gamma_\phi^- t^*)$. We
note that this asymptotic regime satisfies the strong coincidence condition
with $\Omega_{\phi s}=1/(1+r_s^-)$ and $\Omega_{ms}=r_s^-/(1+r_s^-)$. To
leading order the potential $V(\phi)$ becomes

\be
\n{paf}
V(\phi) \simeq\frac{3(2-\gamma_\phi){H^*}^2}
{2\left(1+r_s^-\right)}
\mbox {exp}\left[-\sqrt{\frac{3(1+r_s^-)}
{\gamma_\phi}}\gamma_\phi^-\,\left(\phi-\phi^*\right)\right],
\ee

\noindent   This reproduces the results
of Ref. \cite{zpc}.

\section{Dissipative effects}

The model investigated in the previous section, where the quintessence field
interacts with matter that behaves as a perfect fluid, exhibits a number of
interesting features. However, because of the constraint $r_s^+r_s^-=1$, its
domain of applicability is limited to the evolution of the density ratio
within the interval $1/r_s^-\ge r\ge r_s^-$. Assuming that $r_s^-\simeq
r_0\simeq 0.5$, it implies the upper limit $r\alt 2$. The effect of a scaling
field on CMB anisotropies has been estimated in Ref. \cite{mel} using data
from Boomerang and DASI, providing the constraint $\Omega_\phi\le 0.39$ at
$2\sigma$ during the radiation dominated era. It implies $r>1.6$ at $z\simeq
10^3$ so that $r\simeq 2$ cannot have occurred earlier than
$z\simeq 10^4$. We note however that $r_s^+$ corresponds to infinite redshift
for perfect fluid matter. We will see in this section that a sufficiently
large bulk dissipative pressure in the dark matter fluid allows to shift the
startup redshift at much higher values.

Another line of evidence pointing to dissipative effects in dark matter comes
from the discrepancies between numerical simulations of non-interactive CDM
halo models with observations at the galactic scale \cite{cdm_problems_1},
\cite{cdm_problems_2}. The main discrepancies are the substructure problem,
related to excess clustering on sub-galactic scales, and the cusp problem,
characterized by excessively concentrated cores \cite{halos}\cite{nbody_1},
\cite{nbody_2}, \cite{nbody_3}. Confirmation of these problems would imply
that structure formation is somehow suppressed on small scales. To deal with
them, some kind of self-interaction has been proposed either in cold dark
matter (CDM) models \cite{scdm_1}, \cite{scdm_2}, \cite{scdm_3},
\cite{scdm_4}, \cite{scdm_5} \cite{Lin00} \cite{self}, \cite{Goodman00}
\cite{ann} \cite{Cen}, or in warm dark matter (WDM) models
\cite{wdm_1}-\cite{wdm_6} \cite{suss02} . It is quite reasonable to expect
that dark matter is out of thermodynamical equilibrium and these same
interactions are at the origin of a cosmological dissipative pressure or
thermal effects. A simple estimation shows that a cross section of the order
of magnitude proposed in these halo formation scenarios, corresponding to a
mean free path in the range $1 \, \mbox{kpc}$ to $1 \, \mbox{ Mpc}$, yields at
cosmological densities a mean free  path a bit lower than the Hubble distance.
Hence a description for interacting dark matter as a dissipative fluid at
cosmological scales seems appropriate \cite{Pavon93}.

We may account for the effect of a bulk dissipative pressure $\pi$ in the
matter fluid by the replacement $p_m\to p_m+\pi$, hence $\gamma_m\rho_m\to
\gamma_m\rho_m+\pi$ in Eqs. (\ref{ceq}), (\ref{overall}) and (\ref{deltam}).
So, the effective baryotropic index of matter becomes

\be
\gamma_{m}^{(e)} = \gamma_{m} + \frac{\pi+\Pi}{\rho_{m}}, \qquad
\label{effectivepi}
\ee

\noindent
This means that an ansatz for $\pi/\rho_m$ as a function of $r$ is needed to
calculate the evolution. Here, we complete the model of the previous section
with the inclusion of a bulk viscosity pressure obeying $\pi=-b^2\rho$,
where $b^2$ is a constant. Accordingly, the roots of the  quadratic equation
$r_s\Gamma(r_s)=0$ become

\be
\n{r0pmb}
r_s^\pm=-1-\frac{b^2}{2c^2}+\frac{\gamma_m-\gamma_\phi}{2c^2}\pm\sqrt\Delta,
\ee
\\
where now
\\
\be
\n{delb}
\Delta=\frac{(\gamma_m-\gamma_\phi)^2}{4c^4}-\left(1+\frac{b^2}{2c^2}\right)
\frac{\gamma_m-\gamma_\phi}{c^2}+\frac{b^4}{4c^4}.
\ee

\noindent Thus we find that the constraint between the stationary density
ratios becomes

\begin{equation} \label{rcons}
r_s^+r_s^-=1+\frac{b^2}{c^2}.
\end{equation}

\noindent
We see that the startup density ratio,
increases with the ratio of viscous to interaction pressures.

Now we consider the question whether the transition from matter dominance to
the coincidence era is compatible with a transition from decelerated to
accelerated expansion. This transition implies  $q>0$ for $r\gg 1$ and
$q<0$ on the late attractor stage. With the help of Eqs. (\ref{dr}),
(\ref{p/rm}) and (\ref{00r}) with $k = 0$ the deceleration parameter
$q=-\ddot{a}a/\dot{a}^{2}$ can be written as

\begin{equation} \label{q}
q=\frac{3}{2}\left(\gamma_\phi^{(e)}+\frac{r\Gamma}{1+r}\right)-1\,.
\end{equation}

\noindent In the present model these acceleration transition constraints
translates into
$q_s^+\equiv q(r_s^+)>0>q(r_s^-)\equiv q_s^-$
for the acceleration parameter in the asymptotic regime where
\\
\begin{equation} \label{qr}
q(r)=\frac{3}{2}\left[\gamma_\phi+c^2\left(1+r\right)+
c^2\frac{\left(r-r_s^-\right)\left(r_s^+-r\right)}{1+r}\right]-1\,.
\end{equation}

\begin{figure}[c]
\includegraphics*[scale=.8]{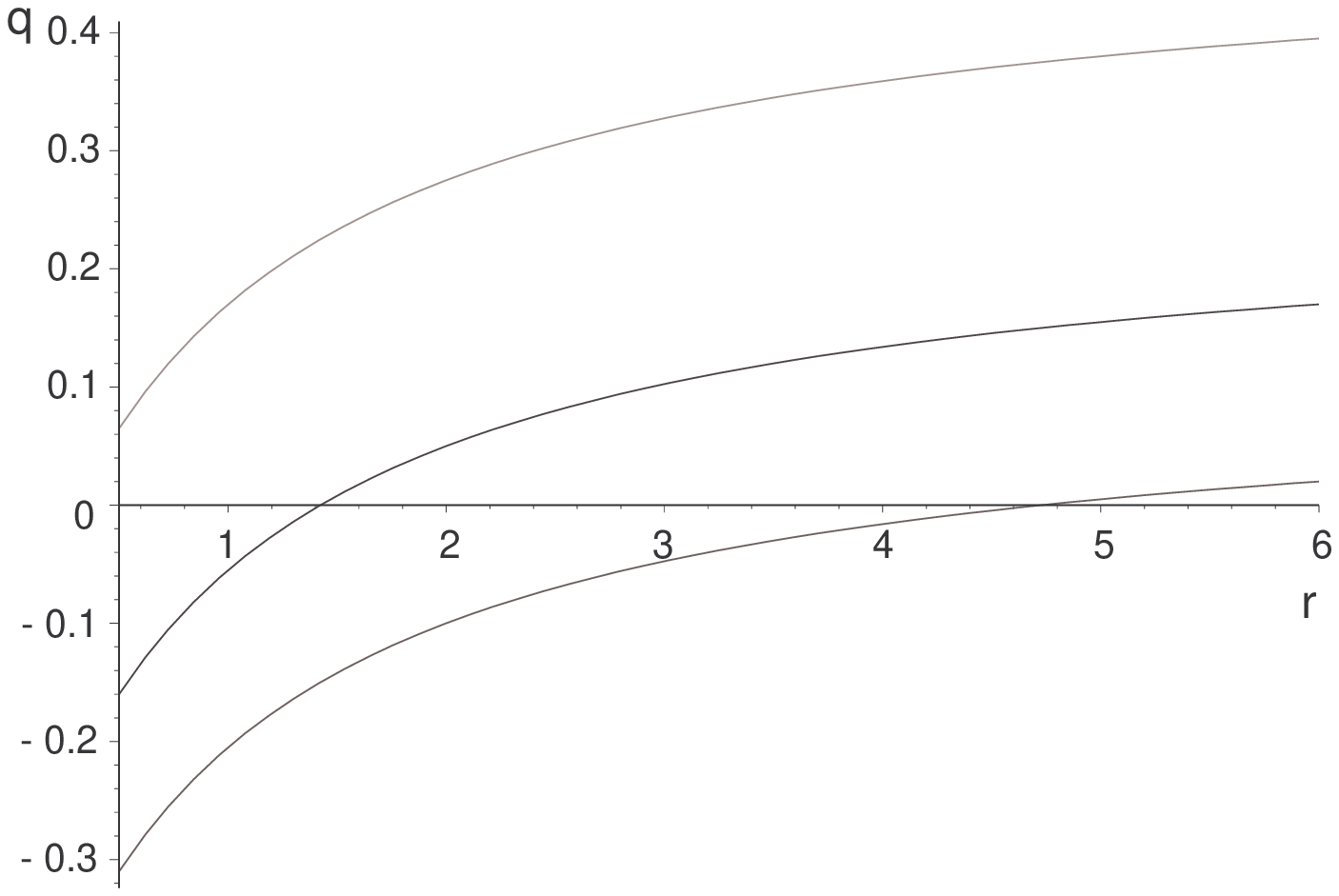}

\caption{\label{fig:q1}
Selected curves of the deceleration parameter $q$ vs the density ratio $r$
between late time ratio $r_s^-=0.5$ and the early time ratio $r_s^+=6$. From
top to bottom, the curves correspond to quintessence baryotropic index $\gamma_\phi=0.65$, $0.5$, and $0.4$.
}

\end{figure}

\noindent
Its derivative
\\
\begin{equation} \label{q'}
q'(r)=\frac{3c^2}{2(1+r)^2}\left(1+r_s^+\right)\left(1+r_s^-\right)
\end{equation}
\\
\noindent
is positive--definite so that the deceleration parameter decreases
monotonically as the Universe expands (see Fig. \ref{fig:q1}).
Then we find that
\\
\begin{equation} \label{qspm}
q_s^\pm=\frac{3}{2}\gamma_\phi^\pm-1 \,,
\end{equation}
\\
and we may write these constraints as
$\gamma_\phi^-<2/3<\gamma_\phi^+$, or equivalently

\begin{equation} \label{consc}
\frac{2/3-\gamma_\phi}{1+r_s^+}<c^2<\frac{2/3-\gamma_\phi}{1+r_s^-}.
\end{equation}

\noindent
Using Eq. (\ref{r0pmb}) we get

\begin{equation} \label{gf1}
\gamma_\phi+\left(1+r_s^+\right)\left(1+r_s^-\right)c^2=\gamma_m \, ,
\end{equation}

\noindent
that combined with (\ref{qspm}) yields

\begin{equation} \label{c2r}
c^2=\frac{3\gamma_m-2-2q_s^-}{3r_s^+\left(1+r_s^-\right)} > 0.
\end{equation}

\noindent
Inserting (\ref{c2r}) into (\ref{consc}) and using (\ref{rcons}), we find that
an accelerating transition implies an upper bound on $r_s^-$

\begin{equation} \label{rs-max}
r_s^-<\frac{1}{3}\left[\left(4+3\frac{b^2}{c^2}\right)^{1/2}-1\right]
\equiv r_{s\mathrm{max}}^-
\end{equation}

\noindent
and a lower bound on $r_s^+$

\begin{equation} \label{rs+min}
r_s^+>\frac{3\left(1+b^2/c^2\right)}{\left(4+3b^2/c^2\right)^{1/2}-1}
\equiv r_{s\mathrm{min}}^+\,.
\end{equation}

\begin{figure}[c]
\includegraphics*[scale=.8]{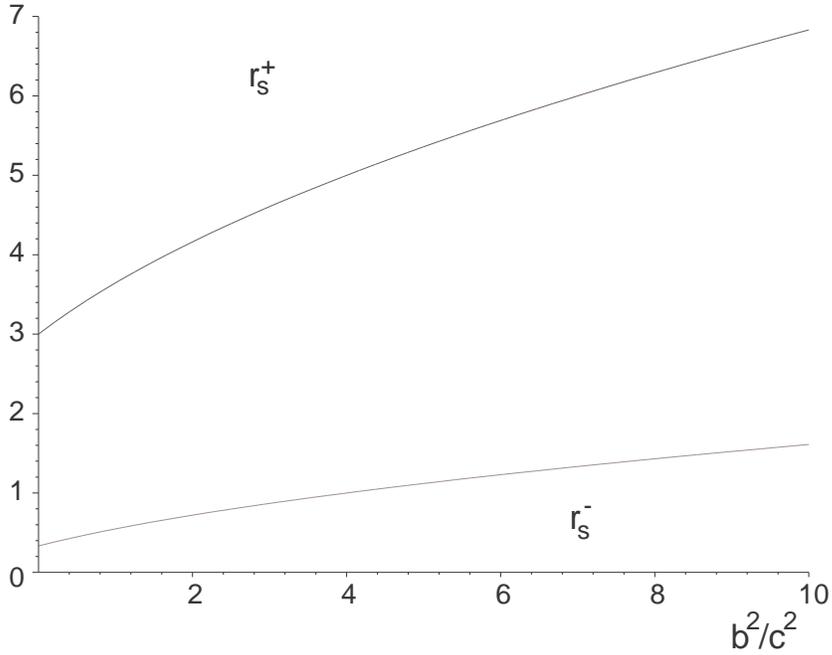}

\caption{\label{fig:rs}
Feasible regions for the late time density ratio $r_s^-$ and the early time
ratio $r_s^+$ for a  transition from decelerated to accelerated expansion to
occur, as a function of the ratio of viscous to interaction pressures
$b^2/c^2$. }

\end{figure}

\noindent We note that both bounds grow with the ratio $b^2/c^2$ and their
values for a perfect fluid (i.e., $b^{2} = 0$) are $1/3$ and $3$ respectively
(see Fig. \ref{fig:rs}). The upper bound (\ref{rs-max}) holds up to the critical
ratio $(b^2/c^2)_c=3r_0^2+2r_0-1$ where $r_{s\mathrm{max}}^-=r_0$. In the high
viscosity regime, above $(b^2/c^2)_c$, the upper bound becomes $r_0$. We also
note that the parameter $\lambda$ has the lower bound
$\lambda_\mathrm{min}=3c^2(r_{s\mathrm{min}}^+ - r_{s\mathrm{max}}^-)$ for
$b^2/c^2\le (b^2/c^2)_c$, that grows with $b^2/c^2$ and has a perfect fluid
value of $8c^2$. On the other hand
$\lambda_\mathrm{min}=3c^2(r_{s\mathrm{min}}^+ - r_0)$, for $b^2/c^2>
(b^2/c^2)_c$.

Further, using Eqs. (\ref{r0pmb}) and (\ref{qspm}), the requirement of
late time acceleration $q_s^- < 0$ becomes

\begin{equation} \label{gammafM}
\gamma_\phi<\frac{1}{3}
\frac{6\gamma_m-4-9c^2\gamma_m-6b^2}{3\gamma_m-2-3c^2-3b^2}\,.
\end{equation}

\noindent Then, the positive--definite character of the quintessence
potential, hence of $\gamma_\phi$, implies that the feasible region in
parameter space $(b^2,c^2)$ has the upper bound
$c^2=2/3-4/(9\gamma_m)-2b^2/(3\gamma_m)$. For CDM it reads $c^2<2/9$ and
$b^2<1/3$.

The density ratio at the beginning of the accelerated expansion
$r_{\mathrm{ac}}$ is given as the root of $q(r_{\mathrm{ac}})=0$. Using Eq.
(\ref{q}) we find

\begin{equation} \label{rac}
r_{\mathrm{ac}}=-\frac{\gamma_\phi-b^2-2/3}{\gamma_m-b^2-2/3}.
\end{equation}

\noindent We see that $r_{\mathrm{ac}}$ also grows with the increase of the
bulk dissipative pressure, so that
$r_{\mathrm{ac}}\ge(2/3-\gamma_\phi)/(\gamma_m-2/3)$ ($>2-3\gamma_\phi$ for
CDM) with the constraint $b^2<\gamma_m-2/3$ ($<1/3$ for CDM).

\begin{figure}[c]
\includegraphics*[scale=.8]{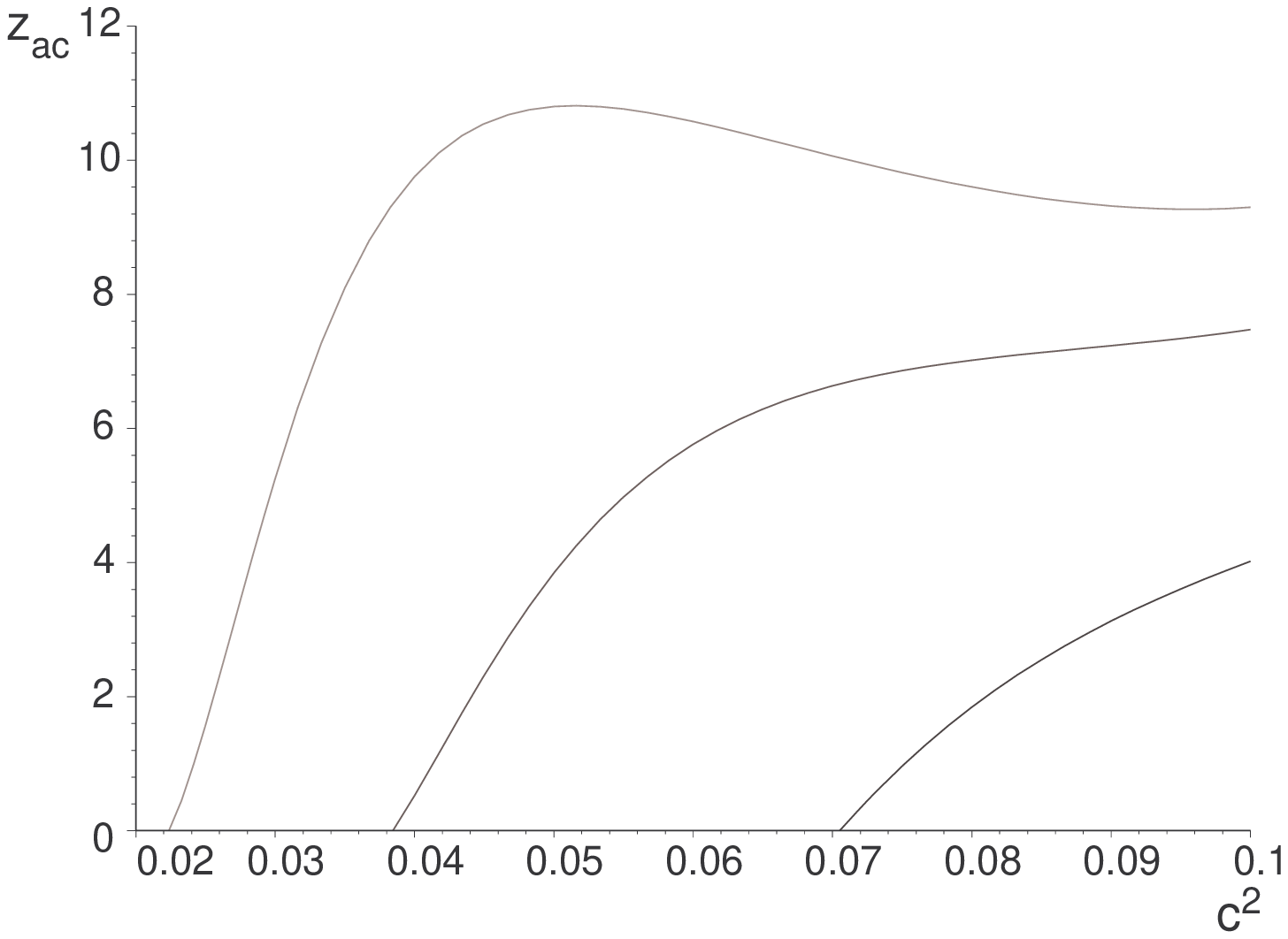}

\caption{\label{fig:za1}
Selected curves of the redshift at the begining of the accelerated expansion
$z_{\mathrm{ac}}$ vs the interaction parameter $c^2$. They correspond to
current density ratio $r_0=0.56$, late time ratio $r_s^-=0.5$, early time
ratio $r_s^+=6$, viscous to interaction presure ratio $b^2/c^2=2$. From top to
bottom, the curves correspond to quintessence baryotropic index  $\gamma_\phi=0.55$, $0.6$,
and $0.7$. }

\end{figure}

\begin{figure}[c]
\includegraphics*[scale=.8]{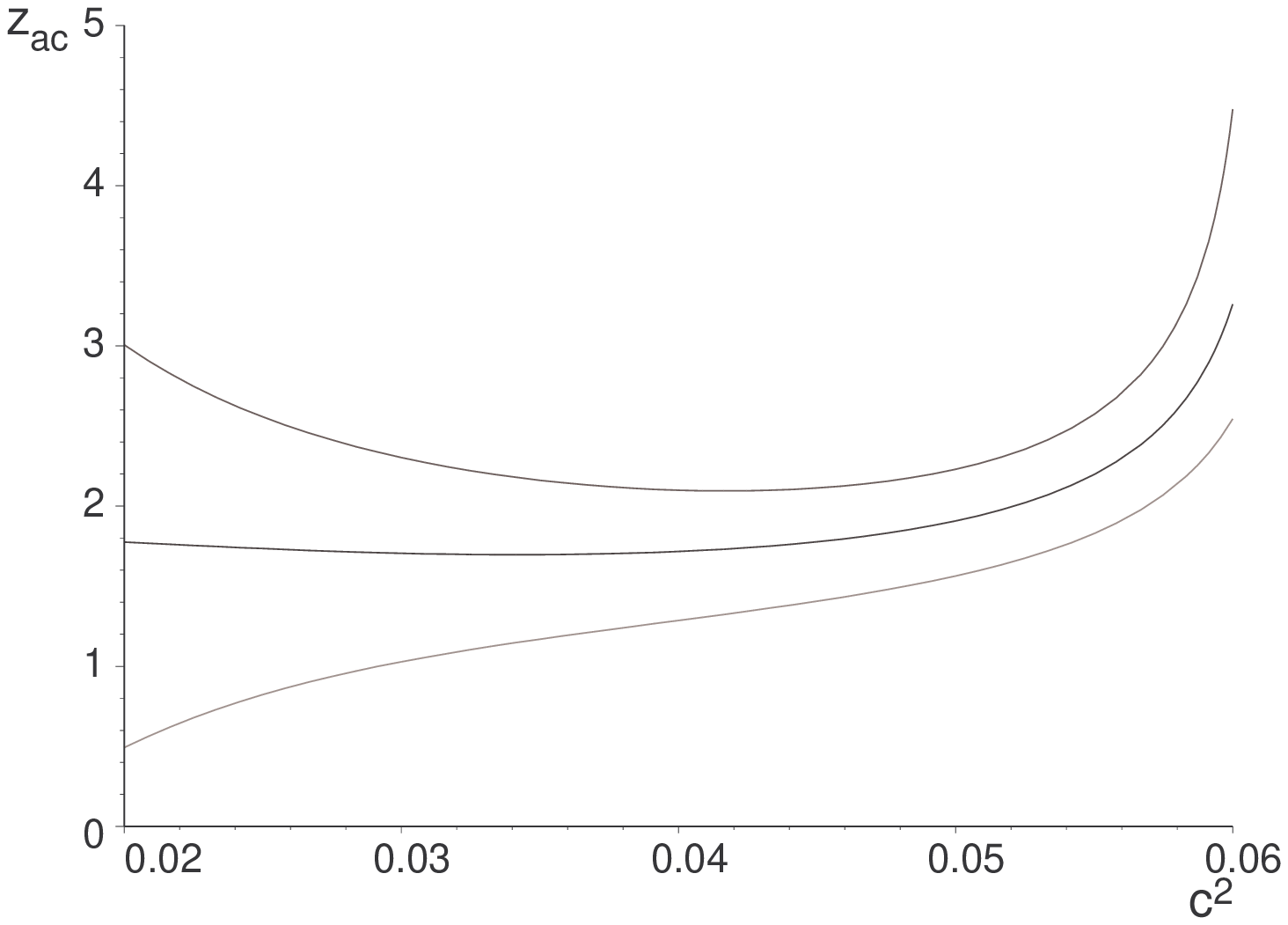}

\caption{\label{fig:za2}
Selected curves of the redshift at the begining of the accelerated expansion
$z_{\mathrm{ac}}$ vs the interaction parameter $c^2$. They correspond to current
density ratio $r_0=0.56$, late time ratio $r_s^-=0.5$, early time ratio
$r_s^+=12$, viscous to interaction presure ratio $b^2/c^2=5$. From top to
bottom, the curves correspond to quintessence baryotropic index  $\gamma_\phi=0.4$, $0.5$, and $0.6$.
}

\end{figure}

Likewise, the inequalities $r_s^+>r_{\mathrm{ac}}>r_0>r_s^-$
must hold. The corresponding redshift $z_{\mathrm{ac}}$ is given by
\\
\begin{equation} \label{zac}
1+z_{\mathrm{ac}}=
\left[
\frac{\left(r_{\mathrm{ac}}-r_s^-\right)\left(r_s^+-r_0\right)}
{\left(r_s^+-r_{\mathrm{ac}}\right)\left(r_0-r_s^-\right)}
\right]^{1/\lambda},
\end{equation}
\\
where we have used Eq. (\ref{xz}). We note that the value of the
acceleration redshift is model dependent. For $\Lambda$CDM models it has been
shown to be close to unity \cite{TurnerRiess}, while it has been argued that
coupling between dark energy and dark matter allows for
$z_{\mathrm{ac}}\simeq 5$
\cite{Amendola02}. We have found that this model leads
either to $z_{\mathrm{ac}}\alt 1$ or much larger values, depending on the
sector of the parameter space (see Figs. \ref{fig:za1} and \ref{fig:za2}).

\section{Observational constraints}
It seems that supernovae of type Ia (SNeIa) may be used as standard candles.
Properly corrected, the difference in their apparent magnitudes is related to
the cosmological parameters. Confrontation of cosmological models to recent
observations of high redshift supernovae ($z\alt 1$) have shown a good fit in
regions of the parameter space compatible with an accelerated expansion
\cite{pedagogical,Perlmutter98,Riess98,Efstat,Perlmutter99,Wang99}. We note,
however, that models like $\Lambda$CDM and QCDM usually require fine tuning to
account for the observed ratio between dark energy and clustered matter, while
QIM models simultaneously provide a late accelerated expansion and solve
the coincidence problem.

Ignoring gravitational lensing effects, the predicted magnitude for an object
at redshift $z$ in a spatially flat homogeneous and isotropic universe is
given by \cite{Peebles93}

\begin{equation}
m(z) =  {\cal M} + 5\log{\cal D}_L(z),
\end{equation}

\noindent
where ${\cal M}$ is its Hubble radius free absolute magnitude and
${\cal D}_L$ is the  luminosity distance in units of
the Hubble radius,

\begin{equation} \label{DL}
{\cal D}_L=\left(1+z\right)\int^z_0 dz'\frac{H_0}{H\left(z'\right)}.
\end{equation}

For noninteracting QCDM models ${\cal D}_L$ can be expressed functionally in
terms of $\gamma_\phi$ (assuming that $\gamma_m$ is a constant), so that the
history of this index $\gamma_\phi(z)$ could in principle be reconstructed
from the magnitude-redshift data of SNeIa alone. Further, using the
conservation equation of the field, the history $V(z)$ of the quintessence
potential could also be reconstructed. For this reason, many authors have
dealt with the recent evolution of $\gamma_\phi$.

When quintessence interacts with matter however, the quintessence baryotropic
index looses this preeminent role. To see why it is necessary to plug the
expansion rate $H(z)$ in Eq. (\ref{DL}). We first note that Eq. (\ref{dr})
can be written as $d\ln r/d\ln(1+z)=3\Gamma$.
Then integrating Eq. (\ref{p/rm}) and inserting in Eq. (\ref{00r}) (for $k=0$)
we find
\\
\begin{equation} \label{Hz1}
\frac{H(z)}{H_0}=\left[\frac{1+r(z)}{1+r_0}\right]^{1/2}
\exp\left[\frac{3}{2}\int_0^z\frac{dz'}{1+z'}
\frac{\gamma_\phi^{(e)}(z')}{\Gamma(z')}\right],
\end{equation}
\\
\noindent
where the density ratio in terms of the redshift is given by

\begin{equation} \label{rz}
r(z)=r_0\exp\left[3\int_0^z\frac{dz'}{1+z'}\Gamma(z')\right]
\end{equation}

\noindent So, ${\cal D}_L$ besides being a functional of the quintessence
baryotropic index also becomes a functional of the interaction and
dissipative pressures through the effective baryotropic indices. As the
histories $\Pi(z)$ and $\pi(z)$  cannot be disentangled from $\gamma_\phi(z)$
in (\ref{Hz1}), the magnitude-redshift data alone cannot reconstruct
$\gamma_\phi(z)$ even in principle when interactions occur.

In virtue of Eqs. (\ref{Hz}) and (\ref{sigma}) we obtain

\begin{equation} \label{DLz}
{\cal D}_L=\left(1+z\right)\left(1+\sigma\right)^{1/2}
\int^z_0\left(1+z'\right)^{-3\gamma_\phi^-/2}
\left[1+\sigma\left(1+z'\right)^\lambda\right]^{-1/2}.
\end{equation}
\\
This integral can be expressed in terms of the hypergeometric function

$$
{\cal D}_L=\left(1+z\right)\frac{\left(1+\sigma\right)^{1/2}}
{\lambda\left(R+1\right)}\left[\left(1+z\right)^{R+1}
{}_2F_1\left(\frac{1}{2},R+1,R+2;-\sigma\left(1+z\right)^\lambda\right)\right.
$$
\begin{equation} \label{DLH}
-\left.{}_2F_1\left(\frac{1}{2},R+1,R+2;-\sigma\right)\right],
\end{equation}

\noindent
where $R=(1-\lambda-3\gamma_\phi^-/2)/\lambda$.
We have used the sample of $38$ high redshift ($0.18 \le z \le 0.83$)
supernovae of Ref. \cite{Perlmutter98}, supplemented with $16$ low redshift
($z < 0.1$) supernovae from the Cal\'an/Tololo Supernova Survey
\cite{Hamuy}. This is described as the ``primary fit'' or fit C in Ref.
\cite{Perlmutter98}, where, for each supernova, its redshift $z_i$, the
corrected magnitude $m_i$ and its dispersion $\sigma_i$ were computed.
We have determined the optimum fit of the QIM model by minimizing a
$\chi^2$ function

\begin{equation} \label{chi4}
\chi^2=\sum^N_{i=1}\frac{\left[m_i-m(z_i;\lambda,R,\sigma,{\cal M})\right]^2}
{\sigma_i^2} \,,
\end{equation}

\noindent
where $N=54$ for this data set. This fit yields $\sigma\simeq 0$
as the most probable value. We note that $\sigma=0$ means that the Universe
is settled at the asymptotic state $r=r_s^-$, with a
constant deceleration parameter given by Eq. (\ref{qspm}).

In other words, due to the limitations of the magnitude-redshift method (see
Ref. \cite{Maor} and references therein) the currently available set of
supernovae located at $z<1$ is unable to provide a clear signature of the
turnover from  matter dominated decelerated phase to the current accelerated
one. Hopefully future observations of type Ia supernovae, such as expected
from the proposed SNAP satellite \cite{SNAP}, combined with other cosmological
observations, will provide much severe constraints on the parameters and
produce a reliable  reconstruction of the evolution of the density ratio.
So, for the purpose of fitting to this set of observations, we may use the
attractor solution $r_s^-$. Then Eq. (\ref{DLH}) simplifies to
\cite{powerlaw,qsa}

\begin{equation} \label{dLsa}
{\cal D}_L(z)=\frac{\left(1+z\right)\left[(1+z)^\beta-1\right]}{\beta } \,,
\end{equation}

\noindent
where $\beta=\lambda(R+1)=-q_s^-=1-1/\alpha$ is minus the
asymptotic deceleration parameter. So $\beta$ increases with $\alpha$,
and $1<\alpha<\infty$ corresponds to $0<\beta<1$. On the attractor
we have

\be
\n{saa}
\gamma_{s}={2\over 3\alpha}=\frac{2}{3}(1-\beta) \,.
\ee

The optimum fit of this attractor model is given by minimizing the
$\chi^2$ function

\begin{equation} \label{chi2}
\chi^2=\sum^N_{i=1}\frac{\left[m_i-m(z_i;\beta,{\cal M})\right]^2}
{\sigma_i^2}.
\end{equation}

\noindent
The most likely values of these parameters are found to be
$(\beta,{\cal M})=(0.395,23.96)$, yielding
$\chi^2_\mathrm{min}/N_{DF}=1.12$ ($N_{DF}=52$), and a
goodness--of--fit $P(\chi^2\ge \chi^2_\mathrm{min})=0.253$.  We
estimate the probability density distribution of the parameters by
evaluation of the normalized likelihood \cite{Lupton93}

\begin{equation} \label{pbetaM}
p(\beta,{\cal M})=\frac{\exp\left(-\chi^2/2\right)}
{\int d\beta \int d{\cal M}\exp\left(-\chi^2/2\right)} \,.
\end{equation}

\begin{figure}[c]
\includegraphics*[scale=.6]{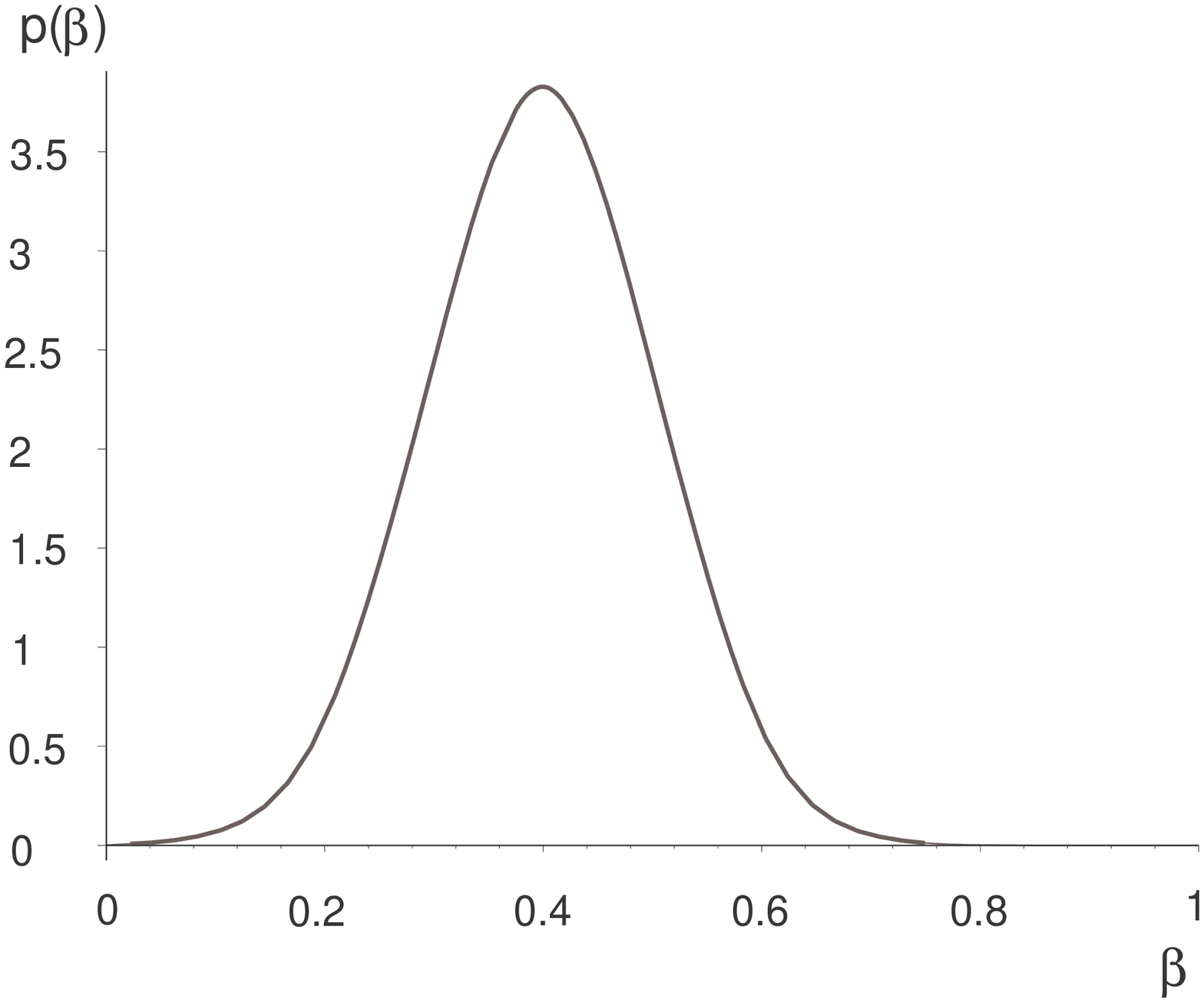}

\caption{\label{fig:pb}
The estimated probability density distribution (normalized likelihood)
for the acceleration parameter $\beta=-q_s^-$ of the asymptotically stable
stationary solution $r=r_s^-$.
}

\end{figure}

\noindent
Then we obtain the probability density distribution for $\beta$
marginalizing $p(\beta,{\cal M})$ over ${\cal M}$. This probability density
distribution $p(\beta)$ is plotted in Fig. \ref{fig:pb} and it yields $\beta=0.398\pm
0.104$ $(1\sigma)$. Hence it can be established that $0.085<\beta<0.711$ with
a confidence level of $0.997$; an accelerated superattractor QIM
universe is strongly supported by this data set, in agreement with a similar
analysis of $\Lambda$CDM and QCDM models
\cite{Perlmutter98,Riess98,Garnavich98,Perlmutter99}.

Stronger bounds on the parameters $\gamma_\phi$, $b^2$ and $c^2$ than those
obtained in the previous section may be obtained from $\gamma_m$ and estimates
of $r_s^\pm$ and $q_s^-$.
The equations to be used are (\ref{qspm}), (\ref{rcons}) and

\begin{equation} \label{cb}
c^2+b^2=\frac{r_s^-}{1+r_s^-}
\left[\gamma_m-\frac{2}{3}\left(q_s^-+1\right)\right]
\end{equation}

\noindent
obtained by combining Eqs. (\ref{r0pmb}) and (\ref{qspm}).  For the
estimates we take as before $r_s^-\simeq r_0$, $q_s^-\simeq -\beta$.
Besides, Big Bang nucleosynthesis and the fluctuations imprinted on
the Cosmic Microwave Background open new windows on the evolution of
the quintessence component \cite{FJ,mel,kneller}. Excluding
quintessence inflation models (e.g. \cite{PV}), primordial
nucleosynthesis is probably the earliest epoch from where we can get
some information about the density ratio. Hence we approximate the
initial density ratio $r_s^+$ as the ratio at the start up of
primordial nucleosynthesis $r_N$ (see the Appendix for further
discussion). The quintessence component, evolving under an
approximately exponential potential, appears in the early Universe as
a form of radiation affecting nucleosynthesis abundance yields and
the heights of the acoustic peaks in the cosmic microwave background
radiation. When decoupled from matter it behaves like a
collisionless, isotropic, and nearly non-clustering component
\cite{FJ,dr}. Such a non--standard component alters the cooling rate
and the upper bound on the change of relativistic energy density is
parameterized in terms of the maximum variation in the effective
number of neutrino species as $\Delta g_*(T_N)<7\Delta
N_{\mathrm{max}}/4$, where $g_*(T_N)=10.75$ in the standard model
with three massless neutrinos.  Then, assuming that the
quintessence--matter interaction switches on after nucleosynthesis,
we get the bound

\begin{equation} \label{rN}
r_s^+\simeq r_N>\frac{4g_*(T_N)}{7\Delta N_{\mathrm{max}}}\simeq 6.14.
\end{equation}

Combining it with Eq. (\ref{c2r}) and taking into account that $\Delta
N_{\mathrm{max}}\simeq 1$ \cite{dolgov}  we get for pressureless matter the
upper bound on the interaction coefficient

\begin{equation} \label{c2<}
c^2<\frac{7\left(1-2q_0\right)\Delta N_{\mathrm{max}}}
{12\left(1+r_0\right)g_*(T_N)}\simeq 0.063.
\end{equation}

Then, inserting this value in Eq. (\ref{qspm}) it follows that
$\gamma_\phi>0.3$. In other words, a cosmological constant is excluded
in this model. Similarly, from Eq.(\ref{cb}) we get
\\
\begin{equation} \label{cbsim}
c^2+b^2
\simeq \frac{r_0}{3\left(1+r_0\right)}\left(1-2q_0\right)\simeq 0.215.
\end{equation}
\\
Besides, assuming a postnucleosynthesis switch on of the interaction, we find
from Eq. (\ref{rcons})
\\
\begin{equation} \label{rconssim}
\frac{b^2}{c^2}\simeq r_Nr_0-1>2.43.
\end{equation}
\\
Hence $b^2$ must lie in the range $(0.152, 0.215)$ and $c^2$
in the interval $(0, 0.063)$. Note that the lower bound (\ref{rconssim}) is
above the critical value $(b^2/c^2)_c=1.07\pm0.38$, suggesting that viscous
effects are important.

\section{Concluding remarks}
We have presented a spatially homogeneous and isotropic interacting
quintessence model that evolves towards a phase of accelerated
expansion and simultaneously solves the coincidence problem. It
provides a reasonable explanation to the embarrassing question, ``why
the contributions of dark matter and dark energy (which in principle
scale at different rates with expansion) to the overall energy
density are of the same order precisely today?'' Rather than
postulating a potential for the quintessence field and specifying its
interaction with the dark matter component, we derived these
quantities from the strong coincidence condition (i.e., that the
density parameters of these two components tend to constant values at
late time).  This  requirement led us to the stationary condition
(Eq.(\ref{stationary})), first obtained in Ref. \cite{enlarged}, as
well as to conclude that the FLRW metric must be spatially flat.

The ratio $r$ between the energy densities is seen to evolve from an initial
unstable value $r_{s}^{+}$ up to the lower and stable asymptotic value
$r_{s}^{-}$ at late time. In terms of these quantities we have introduced the
parameter $\sigma$ that assess how near (or far away) from the asymptotic
state of constant acceleration -see Eq. (\ref{sigma})-  our Universe lies.
The available data seem to suggest that our Universe is close to the
stationary era. On the other hand, they are not sufficient to discriminate
our model from the $\Lambda$CDM model.
Hopefully, the SNAP satellite will provide us with a wealth of high redshift
data likely enough to break the degeneracy and infer the redshift at which
the Universe began accelerating its expansion.

We note that the quintessence--matter interaction and the dissipative pressure
terms imply that the magnitude--redshift relationship is not enough, even in
principle, to reconstruct the evolution of these quantities together with
the quintessence baryotropic index history. Further independent observational
tests are needed for this reconstruction program. Finally, the evolution of
cosmological perturbations predicted by this model remains to be studied.
This will be considered elsewhere.


\begin{acknowledgments}

D.P. and W.Z. acknowledge partial support by the NATO grant PST:CLG.977973,
This work was also partially supported by the University of Buenos Aires under
Project X223 and the Spanish Ministry of Science and Technology under grant
BFM 2000-C-03-01 and 2000-1322.

\end{acknowledgments}

\appendix*
\section{}

Here we collect some complementary considerations regarding our
model.
We begin by noting that we consider a universal coupling of the quintessence
field to all sorts of matter, either baryonic or not. As our main concern is
the investigation of the late universe and a dynamical solution to the
coincidence problem, we consider along this stage a simple two-component
model: matter (excluding radiation) and quintessence (an additional radiation
(relativistic) component would be dynamically irrelevant).

The coupling between matter and quintessence manifests itself as the
nonconservation of their partial stress--energy tensors $\nabla_\mu
T_{(m)}{}^{\mu}_{\nu}=-\nabla_\mu T_{(\phi)}{}^{\mu}_{\nu}\neq 0$.
For the investigation of the dynamics of our homogeneous model, it
suffices to specify the projection of this nonconservation equation
along the velocity of the whole (comoving) fluid  $u^\nu$, that for a
perfect fluid is

\begin{equation} \label{uDTm}
u^\nu\nabla_\mu T_{(m)}{}^\mu_\nu=
-u^\mu\nabla_\mu\rho_m-\Theta (\rho_m+p_m)=-\delta
\end{equation}

\noindent
where, for the specific model we have investigated in Sects. III to V, the
coupling is $\delta = c^2 \rho \Theta $ with $\Theta=\nabla_\alpha u^\alpha$
the expansion scalar and $\rho=u^\mu u^\nu T_{\mu\nu}$, where $T_{\mu \nu}$
denotes the total stress energy tensor.

On the other hand, for the investigation of inhomogeneous perturbations of
this model, it will be necessary to take into account that the velocity of the
components $u_{(m)}^\mu$ and $u_{(\phi)}^\mu$ are different, in general,
from the velocity the overall fluid. Also, these fluids may experience acceleration due
to pressure gradients or due to their coupling (anomalous acceleration).
Perhaps the simplest generalization of (\ref{uDTm}) to this wider framework
is the ``longitudinal coupling''

\begin{equation} \label{DTm}
\nabla_\mu T_{(m)}{}^\mu_\nu=  u_{(m) \nu} \delta .
\end{equation}

\noindent It involves energy transfer between matter and quintessence
with no momentum transfer to matter, so that no anomalous acceleration arises.
Hence this choice is not affected by observational bounds to a
``fifth'' force exerted on the baryons. Clearly other generalizations
of Eq.  (\ref{uDTm}) could be considered that do involve an anomalous
acceleration in matter due its coupling to quintessence. In this
regard we note that because of the universal nature of this coupling,
it could not be detected by differential acceleration experiments. We
also note that the coupling we have proposed is purely
phenomenological and the validity of the expression for $\delta$ is
restricted to cosmological scales (as it depends on magnitudes that
are only well defined in that setting). This means that the form of
the coupling at smaller scales remains unspecified, and the
requirements for the different couplings that could have a
manifestation at these scales are that they give the same (averaged)
coupling $\delta$ at cosmological scales and meet the observational
bounds from the ``local'' experiments \cite{clifford}.

The longitudinal coupling is also a very attractive choice to extend our model
to the radiation era, when the dominating matter component is relativistic, as
it does not involve deviation of relativistic particles from their geodesics.
Indeed the exact solution (\ref{rg}) (valid only for constant $\gamma_m$),
holds for the matter-quintessence domination era -with
$\gamma_m=1$ in the case of CDM- as well as for the radiation domination era
-with $\gamma_m=4/3$. In the usual approximation that $\gamma_m$ drops
instantaneously at equality time $t_E$, all we need to extend our model to the
radiation era is to match both solutions of Eq. (\ref{dr}) at $t_E$ (or
equivalently at $x_E$). Clearly $H$, $r$, the energy densities, $\pi$ and
$\Pi$ are continuous through this transition, and we can also assume  the same
$\gamma_\phi$ for both eras. Hence there is a jump in the slope $\dot r$ at
$t_E$ given by

\begin{equation} \label{[ddr]}
\dot r_>-\dot r_<=H_E r_E
\end{equation}

\noindent where $\dot r_>$ ($\dot r_<$) is the limit from the right $t\to
t_E^+$ (from the left $t\to t_E^-$). That is, $r$ falls more steeply in
the radiation era than in the matter era. For the combined solution the
initial state is the radiation era stationary solution $r_s^+$ that is larger
than the $r_s^+$ of the matter era solution extrapolated to the radiation era.
Then, a lower bound for the matter $r_s^+$ is also a lower bound for the
radiation $r_s^+$, and no bound obtained for the parameters is spoiled. In
fact better bounds could be obtained using the combined solution.

The extension back in time of the coupling $\delta$ involves another
nonuniqueness, and there is no reason to assume that it had the same form also
during the early universe. In particular, all couplings containing terms that
become negligible in comparison with $\delta$ at late times (e.g. terms like
$H^n$ with $n>3$) hold the same property of solving the coincidence problem.
And the results of our model hold for this class of models in an approximated,
asymptotic sense.

In our model we are assuming that the coupling was negligible during the
primordial nucleosynthesis. A simple ``generalized'' coupling with this property is

\begin{equation} \label{deltagen}
\delta_{gen}=3c^2H\rho\exp \, \left(-\frac{H}{H_1}\right),
\end{equation}

\noindent
where the parameter $H_1$ may be chosen at will, so that
$\delta_{gen}\simeq \delta$ for $H\ll H_1$ and $\delta_{gen}\simeq 0$ for
$H\gg H_1$. So, by choosing $H_1$ as the Hubble parameter at some suitable
time after nucleosynthesis (e.g. $H_1\sim 10^{-4}\, \mbox{s}^{-1}$),
all our results for the late universe stand while the coupling
essentially vanishes for times prior to nucleosynthesis end.

Another reason to assume an onset of the interaction at some epoch of the
early universe is the high mass of the quintessence field that grows as

\begin{equation} \label{m2t}
m_\phi^2\simeq\frac{V_0 A^2}{2}\left(\frac{t_0}{t}\right)^2
\end{equation}

\noindent for $t\to 0$. This means that the mass $m_\phi$ becomes arbitrary
large for early enough times. However, because of the coupling with matter,
this very heavy particle could decay into  ``dangerous'' particles like
gravitinos, whose decay products would change the baryon--to--photon ratio
required by a successful nucleosynthesis (see eg. \cite{gravitino}).


\begin{thebibliography}{99}

\bibitem{pedagogical}
S. Perlmutter, Int. J. Mod. Phys. A {15 S1B}, 715 (2000).

\bibitem{Perlmutter98}
S. Perlmutter {\em et al.}, Astrophys. J. {\bf 517}, 565 (1999).

\bibitem{Riess98}
A.G. Riess {\em et al.}, Astron. J. {\bf 116}, 1009 (1998).

\bibitem{paris}
P. Brax {\em et al.}, edits. Proceedings of the XVIIIth IAP Colloquium,
``On the nature of dark energy", held in Paris 1-5 July 2002 (in the press).

\bibitem{Efstat}
G. Efstathiou, S.L. Bridle, A.N. Lasenby, M.P. Hobson, and R.S. Ellis,
Mon. Not. R. Astron. Soc. {\bf 303} L47 (1999).

\bibitem{Perlmutter99}
S. Perlmutter, M.S. Turner and M. White,
Phys. Rev. Lett. {\bf 83}, 670 (1999).

\bibitem{Bahcall99}
N.A. Bahcall {\em et al.}, Science {\bf 284}, 1481 (1999).

\bibitem{Efstat02}
G. Efstathiou {\em et al.}, Mon. Not. R. Astron. Soc. {\bf 330}, L29 (2002).

\bibitem{melchiorri}
A. Melchiorri, talk delivered at the XVIIIth IAP Colloquium,
``On the nature of dark energy", held in Paris 1-5 July 2002, edits.
P. Brax {\em et al.} (in the press).

\bibitem{luca}
L. Amendola, Phys. Rev D {\bf 62}, 043511 (2000);
D. Tocchini--Valentini and L. Amendola, Phys. Rev. D {\bf 65}, 063508
(2002).

\bibitem{zpc}
W. Zimdahl, D. Pav\'{o}n and L.P. Chimento, Phys. Lett. B {\bf 521}, 133 (2001);
W. Zimdahl and D. Pav\'{o}n, to be published in Gen. Relativ. Grav. (2003),
preprint {\sf astro-ph/0210484}

\bibitem{enlarged}
L.P. Chimento, A.S. Jakubi and D. Pav\'on, Phys. Rev. D {\bf 62},
063508 (2000).

\bibitem{antf}
W. Zimdahl, D.J. Schwarz, A.B. Balakin and D. Pav\'on,
Phys. Rev. D {\bf 64}, 063501 (2001).

\bibitem{qsa}
Chimento L. P., Jakubi A. S. and Zuccal\'a N.A.,
Phys. Rev. D 63, 103508 (2001).

\bibitem{flat}
P. de Bernardis {\em et al.}, Nature {\bf 404}, 955 (2000);
S. Hanany {\em et al.}, Astrophys. J. {\bf 545}, L5 (2000);
C.B. Netterfield {\em et al.}, astro-ph/0104460;
C. Pryke {\em et al.}, astro-ph/0104490

\bibitem{tytler}
D. Tytler {\em et al.}, Physica Scripta {\bf T85}, 12 (2000);
J.M. O'Meara {\em et al.}, Astrophys. J. {\bf 552}, 718 (2001).

\bibitem{turner2001}
M.S. Turner, astro-ph/0106035.

\bibitem{mel}
R. Bean, S. H. Hansen, and A. Melchiorri
Phys. Rev. D {\bf 64 } 103508 (2001).



\bibitem{cdm_problems_1}  B. Moore, Nature, {\bf 370}, 629, (1994).

\bibitem{cdm_problems_2}  R. Flores and J. P. Primack, Astrophys. J.
{\bf 427}, L1, (1994).


\bibitem{halos}
A. Burkert and J. Silk,
in {\em Dark Matter in Astro and Particle Physics},
edited by H.V. Klapdor-Kleingrothaus and L. Baudis
(IOP, Bristol, 1999).


\bibitem{nbody_1}  J.F. Navarro, C.S. Frenk and S.D.M. White, Astrophys.
J. {\bf 462}, 563, (1996); {\it ibid}  {\bf 490}, 493 (1997).

\bibitem{nbody_2}  B. Moore {\em et al.}, Monthly Not. R. Astr. Soc.
{\bf 310}, 1147, (1999).

\bibitem{nbody_3}  S. Ghigna {\em et al.}, {\sf astro-ph/9910166}.




\bibitem{scdm_1} D.N. Spergel and P.J. Steinhardt, Phys Rev Lett.,
{\bf 84}, 3760, (2000).

\bibitem{scdm_2}  A. Burkert, Astrophys. J. Lett. {\bf 534}, 143, (2000).

\bibitem{scdm_3}  C. Firmani {\em et al}, Monthly Not. R. Astr. Soc.
{\bf 315}, 29, (2000).

\bibitem{scdm_4} R.N. Mohapatra, S. Nussinov, V. L. Teplitz.
e-Print Archive: {\sf hep-ph/0111381}.

\bibitem{scdm_5} Rabindra N. Mohapatra and Vigdor L. Teplitz,
Phys. Rev. D {\bf 62}, 063506 (2000).


\bibitem{Lin00}
W.B. Lin,  D.H. Huang,   X. Zhang, and  R. Brandenberger,
Phys. Rev. Lett. {\bf 86}, 954 (2001).

\bibitem{self}
D.N. Spergel and P.J. Steinhardt,
Phys. Rev. Lett. {\bf 84}, 3760 (2000);
J. P. Ostriker,
{\em ibid.} {\bf 84}, 5258 (2000);
S. Hannestad,
``Galactic halos of self--interacting dark matter,"
astro-ph/9912558;
C. Firmani {\em et al.},
Mon. Not. R. Astron. Soc. {\bf 315}, L29 (2000).

\bibitem{Goodman00}
J. Goodman,
New Astronomy {\bf 5}, 103 (2000).

\bibitem{ann}
M. Kaplinghat, L. Knox, and M.S. Turner,
Phys. Rev. Lett. {\bf 85}, 3335 (2000).

\bibitem{Cen}
R. Cen,
Astrophys. J. Lett. (to be published),
astro-ph/0005206




\bibitem{wdm_1}  S. Colombi, S. Dodelson and L. Widrow, Astrophys. J.
{\bf 458}, 1 (1996).

\bibitem{wdm_2}  R. Schaeffer and J. Silk, {\it ApJ}, {\bf 332}, 1, (1998).

\bibitem{wdm_3}  C.J. Hogan, {\sf astro-ph/9912549}.

\bibitem{wdm_4}  S. Hannestad and R. Scherrer,  Phys. Rev. D {\bf 62},
043522, (2000).

\bibitem{wdm_5}  J.J. Dalcanton and C.J. Hogan, {\sf astro-ph/0004381}.

\bibitem{wdm_6}  C.J. Hogan and J.J. Dalcanton, {\sf astro-ph/0002330}.

\bibitem{suss02}
L.G. Cabral-Rosetti, T. Matos, D. Nu\~nez, and R.A. Sussman,
Class. Quantum Grav. {\bf 19} 3603 (202).

\bibitem{Pavon93}
D. Pav\'on, and W. Zimdahl,
Physics Letters A {\bf 179}, 261 (1993).

\bibitem{TurnerRiess}
M.S. Turner and A.G.Riess,
astro-ph/0106051

\bibitem{Amendola02}
L. Amendola, {\sf astro-ph/0209494}


\bibitem{Wang99}
L. Wang, R. Caldwell, J.P. Ostriker, and P.J. Steinhardt,
Astrophys. J. {\bf 530}, 17 (2000).


\bibitem{Peebles93}
P.J.E. Peebles,
``Principles of Physical Cosmology"
(Princeton University Press, Princeton, New Jersey, 1993).

\bibitem{Hamuy}
M. Hamuy, M.M. Phillips, J. Maza, N.B. Suntzeff, R.A. Schommer, and
R. Avil\'{e}s, Astrophys J. {\bf 112}, 2391 (1996).

\bibitem{Maor}
I. Maor, R. Brustein, J. McMahon, and P.J. Steinhardt
Phys. Rev. D {\bf 65} 123003 (2002).

\bibitem{SNAP}
Levi M.,  et~al., 2000, http://snap.lbl.gov.

\bibitem{powerlaw}
J.A.S. Lima and J.S. Alcaniz,
Astron. Astrophys. {\bf 348} 1 (1999);
M. Kaplinghat, G. Steigman, I. Tkachev, and T.P. Walker
Phys. Rev. D {\bf 59} 043514 (1999).


\bibitem{Lupton93}
R. Lupton,
``Statistics in Theory and Practice"
(Princeton University Press, Princeton, 1993).

\bibitem{Garnavich98}
P.M. Garnavich {\em et al.},
Astrophys. J. {\bf 509}, 74 (1998).

\bibitem{FJ}
P. Ferreira and M. Joyce,
Phys. Rev. D {58}, 023503 (1998).

\bibitem{kneller}
J. P. Kneller and G. Steigman,
{\sf astro-ph/0210500}

\bibitem{PV}
P. J. E. Peebles and A. Vilenkin,
Phys. Rev. D \textbf{59}, 063505 (1999).

\bibitem{dr}
S. Mukohyama,
Phys. Lett. B {473}, 241 (2000).

\bibitem{dolgov}
A.D. Dolgov, ``Big Bang Nucleosynthesis'',{\sf hep-ph/0201107}.

\bibitem{clifford} C.M. Will, ``The confrontation between general
relativity and experiment: A 1998 update", {\sf gr-qc/9811036 };
C.M. Will, Living Rev. Relativity {\bf 4}, 4 (2001).


\bibitem{gravitino}
J. Ellis, J. E. Kim and D. V. Nanopoulos,
Phys. Lett. B \textbf{145}, 181 (1984);
M. Kawasaki and T. Moroi,
Prog. Theor. Phys. \textbf{93}, 879 (1995);
S. Sarkar, Rep. Prog. Phys. \textbf{59}, 1493 (1996);
K. Jedamzik, astro-ph/0112226

\end{thebibliography}
\end{document}